\documentclass[prc,showpacs,nofootinbib,preprint]{revtex4-1}
\usepackage{epsfig,graphics,color}
\usepackage{subfigure}
\usepackage{bbm}
\usepackage{mathtools}
\usepackage{amssymb}
\usepackage{mathrsfs}
\usepackage{amsmath}
\usepackage{amssymb}
\usepackage{amsmath}
\usepackage{pdfsync}
\definecolor{annr}{rgb}{0.86, 0.08, 0.24} 
\definecolor{anng}{rgb}{0.13, 0.55, 0.14}  
\definecolor{annb}{rgb}{0.0, 0.28, 0.67}   
\definecolor{mono}{rgb}{0.65, 0.65, 0.65} 

\usepackage{pifont}

\usepackage{array}
\newcolumntype{L}{>{$}l<{$}}
\newcolumntype{C}{>{$}c<{$}}
\newcolumntype{R}{>{$}r<{$}}

\newcommand{\bra}[1]{\langle #1|}
\newcommand{\ket}[1]{|#1\rangle}

\newcommand{\roundbra}[1]{(#1|}
\newcommand{\roundket}[1]{|#1)}

\def\one{{\bf 1}\,}
\def\quarknumberoperator{{\mathbbm 1}\,}

\def\tr{{\rm tr} \,}

\def\w2{\tilde w^2}
\def\ws2{1}

\newlength{\lslash}

\begin{document}
\title{Constraints from a large-$N_c$ analysis on \\ meson-baryon interactions at chiral order $Q^3$
}
\author{Yonggoo Heo$^1$, C. Kobdaj$^1$\footnote{
    \text{Corresponding authors.}
} and Matthias F.M. Lutz$^{2,3 \,\rm a}$}
\affiliation{$^1$ Suranaree University of Technology, Nakhon Ratchasima, 30000, Thailand}
\affiliation{$^2$ GSI Helmholtzzentrum f\"ur Schwerionenforschung GmbH,\\
  Planck Str. 1, 64291 Darmstadt, Germany}
\affiliation{$^3$ Technische Universit\"at Darmstadt, D-64289 Darmstadt, Germany}
\date{\today}
\begin{abstract}
  We consider the chiral Lagrangian for  baryon fields with $J^P =\frac{1}{2}^+$ or 
  $J^P =\frac{3}{2}^+$ quantum numbers as constructed from QCD with up, down and strange quarks. 
  The specific class of counter terms that are of chiral order $Q^3$ and contribute to meson-baryon interactions at the two-body level is constructed.  Altogether we find 24 terms. In order to pave the 
  way for realistic applications  we establish a set of 22 sum rules for the low-energy constants as they are implied by QCD in the large-$N_c$ limit. Given such a constraint there remain only 2 
  independent unknown parameters that need to be determined by either Lattice QCD simulations or directly from experimental cross section measurements. At subleading order we arrive at 5 parameters. 
\end{abstract}

\pacs{25.20.Dc,24.10.Jv,21.65.+f}
\keywords{Chiral extrapolation, Large-$N_c$, chiral symmetry, flavor $SU(3)$}
\maketitle
\tableofcontents
\newpage

\section{Introduction}

Still after many decades of vigorous studies the outstanding challenge of modern physics is to establish a rigorous link of QCD to low-energy hadron physics as it is observed in the many experimental cross section measurements. After all it is the only fundamental field theory there is 
that leads to the emergence of structure as a consequence of truly non-perturbative interactions in a quantum field theory. On the one-hand the data set is extended recently by LHCb, BES, COMPASS, Belle with more and more exciting new phenomena, on the other hand there is a huge data set on pion and photon induced reactions in the resonance region which still up today is not understood in terms of QCD dynamics \cite{Lutz:2015ejy,Pennington:2016dpj}. Such reactions constitute the doorway of understanding non-perturbative QCD, like studies of the hydrogen atom paved the way of understanding QED.  

While simulations of QCD on finite lattices made considerable progress the last decade it is still not feasible to derive cross sections systematically as measured in the laboratory in the resonance region of QCD. Thus at present it may be of advantage to resort to a well established method 
of modern physics. Derive the implications of the fundamental theory by matching it to effective field theory approaches that are formulated in terms of the relevant degrees of freedom. 

With the great advances of lattice QCD simulations such an approach is going through a revolution at present since the effective field theory can now be scrutinized systematically by QCD lattice data. In turn, the typically quite large set of low-energy constants can be derived from QCD prior to confronting the effective field theory to scattering data taken in the laboratory. This has been emphasized and illustrated in the  recent work \cite{Guo:2018kno}. 
Some results for sets of low-energy constants have already been obtained from the masses of baryons and  mesons in their ground states with $J^P = \frac{1}{2}^+, \frac{3}{2}^+$ and 
$J^P = 0^-, 1^-$ quantum numbers \cite{Lutz:2014oxa,Lutz:2018cqo,Guo:2018kno,Bavontaweepanya:2018yds,Heo:2018qnk}.

Since the majority of available lattice data was taken at unphysical quark masses it is mandatory to establish reliable tools to translate such data back to the physical case. The fact that lattice data are typically for unphysical hadrons so far we see as a fortunate circumstance since this way information on QCD is provided that cannot be inferred from the PDG or any experimental cross section so easily. Moreover, the determination of large sets of low-energy constants  from  lattice data on the hadron ground state  masses at various unphysical quark masses appears to be much easier and better controlled  as compared to their extraction from the first few available phase shifts as computed on QCD lattices at unphysical quark masses. 

Here we wish to emphasize that our strategy how to pave the way towards the understanding of non-perturnative QCD relies heavily on our recent claim that the chiral Lagrangian properly formulated for the physics of up, down and strange quarks, can be successfully applied to low-energy QCD once it is set up in terms of on-shell meson and baryon masses. It was demonstrated that then the size of the physical strange quark mass does not prohibit the application of the chiral Lagrangian. This is contrasted by the conventional $\chi$PT  approach, in which bare masses are to be used inside any loop expression. Here any low-orders application to the flavor SU(3) case should be avoided, being of no physical significance.

The purpose of the current study is to further prepare the quantitative application of the chiral Lagrangian with three light flavors to meson-baryon scattering data. 
Our target is the set of counter terms that carry chiral order $Q^3$ and contribute to meson-baryon scattering at the two-body level. Such $Q^3$ counter terms play a decisive role in the chiral dynamics of the meson-baryon systems. As was pointed out already in \cite{Lutz2002a} only in the presence of such terms it may be feasible to establish a universal set of $Q^2$ counter terms that describe 
pion, kaon and antikaon nucleon scattering data. Though there is a plethora of works \cite{Kaiser:1995eg,Lutz:1997wt,Oset:1997it,Jamin:2000wn,Jido:2003cb,Mai:2009ce,Bruns:2010sv,Bruns:2012eh,Ikeda:2012au,Oller:2013dxa,Feijoo:2015yja,Ramos:2016odk} that fit the $Q^2$ counter terms to pion-nucleon, kaon-nucleon or antikaon-nucleon scattering, the only so far a univeral approach is documented in  \cite{Lutz2002a}. In turn there are various mutually non-compatible sets of the $Q^2$ counter terms available. 

We would argue that there are also still some residual deficiencies in \cite{Lutz2002a} which may hamper the direct use of the most comprehensive set low-energy constants as extracted from the published lattice data set on the baryon octet and decuplet masses in \cite{Lutz:2018cqo}. Most severe, we would argue, are the particularities of the unitarization schemes. Within the flavor SU(3) framework so far all published works rely on neglect or improper treatment of left-hand branch points. Though we do not expect this to lead to huge qualitative issues, a quantitative and controlled study of in particular p-wave phase shifts should consider it in a reliable manner. We feel this to be an achievable request owing to the fact that such a scheme exists by now with \cite{Gasparyan:2010xz,Danilkin:2010xd,Gasparyan:2011yw}. So far it was applied only to the flavor SU(2) case with the $\pi N$ and $\gamma N$ channels. 

Within a flavor SU(3) context such $Q^3$ terms were first used in \cite{Lutz2002a}. Later
the complete order $Q^3$ Lagrangian was constructed in \cite{Frink:2006hx,Oller:2007qd} for the baryon octet fields. To the best knowlege of the authors such counter terms have not been constructed so far involving the baryon decuplet fields. We are aware of the recent work \cite{Jiang:2018mzd}, which, however, provides partial results only. Since we wish to derive sum rules for the $Q^3$ low-energy constants 
from large-$N_c$ QCD \cite{tHooft74,Witten1979} a reliable construction of the latter terms is the target of the first part of our work in section II. It follows the second part with section III in which we apply large-$N_c$ QCD in order to derive sum rules for the set of $Q^3$ low-energy constants. Here we follow the framework previously established in \cite{Luty1994,Dashen1995,Lutz:2010se}. In our case we compute the contributions of the $Q^3$ counter terms to the correlation function with two axial-vector  and one vector current in the baryon ground states. From a study of the latter the desired sum rules will be derived. 

\newpage 

\section{Chiral Lagrangian with baryon octet and decuplet fields} \label{section:chiral-lagrangian}

We recall the conventions for the chiral Lagrangian as used in the current work \cite{Krause1990,Lutz2002a,LutzSemke2010,Lutz:2018cqo}.
The hadronic fields as decomposed into their isospin multiplets are
\begin{eqnarray}
&& \Phi = \tau \cdot \pi (140)
+ \alpha^\dagger \cdot  K (494) +  K^\dagger(494) \cdot \alpha
+ \eta(547)\,\lambda_8\,,
\nonumber\\
&& \sqrt{2}\,B
   =   \alpha^\dag\cdot{N}(939) + \lambda_8\,\Lambda(1115)
  + \vec\tau\cdot\vec\Sigma(1195) + {\Xi}^T(1315) \,i\sigma_2 \cdot\alpha
  \,,\nonumber\\
&& \alpha^\dagger = {\textstyle{1\over \sqrt{2}}}\,(\lambda_4+i\,\lambda_5 ,\lambda_6+i\,\lambda_7 )\,,\qquad \qquad \qquad
\vec\tau = (\lambda_1,\lambda_2, \lambda_3)\,,
\end{eqnarray}
where the matrices $\lambda_i$ are the Gell-Mann generators of the SU(3) algebra.
The numbers in the brackets recall the approximate masses of the particles in units of
MeV. Of central importance is the covariant derivative
\begin{eqnarray}
( D_\mu B)^i_j = \partial_\mu B^i_j + (\Gamma_\mu)^i_h\,B^h_j - B^i_h\,(\Gamma_\mu)^h_j\,,
\label{example-Dmu}
\end{eqnarray}
as introduced in terms of the chiral connection $\Gamma_\mu$.
The chiral connection with $\Gamma_\mu=-\Gamma_\mu^\dagger$  and other convenient chiral building blocks are constructed in terms of the chiral fields $\Phi$ in a non-linear fashion such that the all chiral Ward identities of QCD are 
recovered in systematic applications of the chiral Lagrangian \cite{Gasser1983a,Gasser1983b,Krause1990}. We write
\begin{eqnarray}
&&\Gamma_\mu ={\textstyle{1\over 2}}\,u^\dagger\,\Big[\partial_\mu -i\,(v_\mu + a_\mu) \Big]\,u
+{\textstyle{1\over 2}}\, u\,\Big[\partial_\mu -i\,(v_\mu - a_\mu)\Big]\,u^\dagger \,,
\nonumber\\
&& \textcolor{black}{ U_\mu = {\textstyle{1\over 2}}\,u^\dagger \, \big(
\partial_\mu \,e^{i\,\frac{\Phi}{f}} \big)\, u^\dagger
-{\textstyle{i\over 2}}\,u^\dagger \,(v_\mu+ a_\mu)\, u
+{\textstyle{i\over 2}}\,u \,(v_\mu-a_\mu)\, u^\dagger\;, \qquad \qquad 
 u = e^{i\,\frac{\Phi}{2\,f}} } \,,
\nonumber\\ 
&& H_{\mu\nu} = 
  {D}_\mu\,i\,U_\nu + {D}_\nu \,i\,U_\mu
  \, ,\qquad \qquad \qquad \qquad 
  {D}_{\mu\nu} =  {D}_\mu{D}_\nu + {D}_\nu{D}_\mu \,,
\end{eqnarray}
where we emphasize the presence of the classical vector and axial-vector source fields, $v_\mu$ and $a_\mu$ of QCD \cite{Gasser1983a,Gasser1983b}. The important merit of all building blocks $B, U_\mu, H_{\mu\nu}$ and $D_{\mu \nu}$ lies in their identical chiral transformation properties. Thus, the action of the covariant derivatives is implied by the example case (\ref{example-Dmu}).

As derived first in \cite{Lutz2002a} there are 10 independent symmetry conserving $Q^3$ terms that are needed in the baryon octet sector. Such terms were studied in momentum space properly projected onto the kinematics required in meson-baryon scattering process. Initially there were 20 terms considered. It was shown in \cite{Lutz2002a} that only 10 terms are independent. This result was established by an evaluation of the s- and p-wave projections of their contributions to the scattering amplitudes. Explicit expressions how such terms contribute to the meson-baryon interaction kernel were provided in Appendix B of that work. 

This result was confirmed later in \cite{Frink:2006hx,Oller:2007qd} based on a complementary strategy. In fact, initially the authors of \cite{Oller:2007qd} claimed the relevance of 11 terms in \cite{Oller2006}. A result inconsistent with the original finding in \cite{Lutz2002a}. This error was corrected first in \cite{Frink:2006hx}. In the current work we use the 10 terms in their following representation
\begin{eqnarray}
  && {\mathcal L}^{(3)}_{[8]\,[8]} =- u^{}_{1}\,\tr\,\bar{B}\,\gamma^\mu\,{B}\,[{U}^\nu,\,{H}_{\mu\nu}]_-
  - u^{}_{2}\,\tr\,\bar{B}\,[{U}^\nu,\,{H}_{\mu\nu}]_-\,\gamma^\mu\,{B}
  \nonumber\\
  &&  \qquad \quad   - \,\tfrac{1}{2}\,u^{}_{3}\,\big(
  \tr\,\bar{B}\,{U}^\nu\,\gamma^\mu\,\tr\,{H}_{\mu\nu}\,{B}  + \text{h.c.}
  \big)
  \nonumber\\
  &&  \qquad  \quad - \,\tfrac{1}{2}\,u^{}_{4}\,\big(
  \tr\,\bar{B}\,\gamma^\lambda\,({D}^{\mu\nu}{B})\,[{U}_\lambda,\,{H}_{\mu\nu}]_-
  + \tr\,({D}^{\mu\nu}\bar{B})\,\gamma^\lambda\,{B}\,[{U}_\lambda,\,{H}_{\mu\nu}]_-
  \big)
  \nonumber\\ 
  && \qquad  \quad  - \,\tfrac{1}{2}\,u^{}_{5}\,\big(
  \tr\,\bar{B}\,[{U}_\lambda,\,{H}_{\mu\nu}]_-\,\gamma^\lambda\,({D}^{\mu\nu}{B})
  + \tr\,({D}^{\mu\nu}\bar{B})\,[{U}_\lambda,\,{H}_{\mu\nu}]_-\,\gamma^\lambda\,{B}
  \big)
  \nonumber\\ 
  && \qquad \quad  
  -\, \tfrac{1}{4}\,u^{}_{6}\,\big(
  \tr\,\bar{B}\,{U}_\lambda\,\gamma^\lambda\,\tr\,{H}_{\mu\nu}\,({D}^{\mu\nu}{B})
  +\tr\ ({D}^{\mu\nu}\bar{B})\,{U}_\lambda\,\gamma^\lambda\,\tr\,{H}_{\mu\nu}\,{B}
  + \text{h.c.}
  \big)   
  \nonumber\\
  &&  \qquad \quad 
  -\, \tfrac{1}{2}\,u^{}_{7}\,\big(
  \tr\,\bar{B}\,\sigma^{\lambda \mu}\,({D}^{\nu}{B})\,[{U}_\lambda,\,{H}_{\mu \nu}]_+
  - \tr\,({D}^{\nu} \bar{B})\,\sigma^{\lambda \mu}\,{B}\,[{U}_\lambda,\,{H}_{\mu \nu}]_+
  \big)
  \nonumber\\ 
  && \qquad \quad 
  -\, \tfrac{1}{2}\,u^{}_{8}\,\big(
  \tr\,\bar{B}\,[{U}_\lambda,\,{H}_{\mu \nu}]_+\,\sigma^{\lambda \mu}\,({D}^{\nu}{B})
  - \tr\,({D}^{\nu}\bar{B})\,[{U}_\lambda,\,{H}_{\mu \nu}]_+\,\sigma^{\lambda \mu}\,{B}
  \big)
  \nonumber\\ 
  && \qquad \quad 
  - \,\tfrac{1}{4}\,u^{}_{9}\,\big(
  \tr\,\bar{B}\,{U}_\lambda\,\sigma^{\lambda \mu }\,({D}^{\nu}{B})\,{H}_{\mu \nu}
  - \tr\,({D}^{\nu}\bar{B})\,{U}_\lambda\,\sigma^{\lambda \mu }\,{B}\,{H}_{\mu \nu}
  + \text{h.c.}
  \big)
  \nonumber\\ 
  && \qquad \quad   - \,\tfrac{1}{2}\,u^{}_{10}\,\big(
  \tr\,\bar{B}\,\sigma^{\lambda \mu}\,({D}^{\nu}{B})\,\tr\,{U}_\lambda\,{H}_{\mu \nu}
  - \tr\,({D}^{\nu}\bar{B})\,\sigma^{\lambda \mu}\,{B}\,\tr\,{U}_\lambda\,{H}_{\mu \nu}
  \big) 
  \,.
\end{eqnarray}

We turn to the decuplet sector.  The construction of the chiral Lagrangian is straightforward following the 
rules established by Krause in  \cite{Krause1990}. We use here the conventional Rarita-Schwinger fields to interpolate to the decuplet of the spin-three-half states. The baryon decuplet field $B^{ijk}_\mu$ comes with three fully symmetric flavor indices, $i,j,k = 1,2,3$ as
\begin{align}
  \label{def:field-representation-decuplet}
  \begin{array}{llll}
    B_{\mu}^{111} = \Delta^{++}_\mu
    \,, \qquad \;\;\;&
    B_{\mu}^{112} = \Delta^{+}_\mu/\sqrt3
    \,, \qquad &
    B_{\mu}^{122} = \Delta^{0}_\mu/\sqrt3
    \,, \qquad &
    B_{\mu}^{222} = \Delta^{-}_\mu
    \,,\\
    B_{\mu}^{113} = \Sigma^{+}_\mu/\sqrt3
    \,,&
    B_{\mu}^{123} = \Sigma^{0}_\mu/\sqrt6
    \,,&
    B_{\mu}^{223} = \Sigma^{-}_\mu/\sqrt3
    \,,& \\
    B_{\mu}^{133} = \Xi^{0}_\mu/\sqrt3
    \,,&
    B_{\mu}^{233} = \Xi^{-}_\mu/\sqrt3
    \,,& & \\
    B_{\mu}^{333} = \Omega^{-}_\mu
    \,,& & &
  \end{array}
\end{align}
where the components are identified with the states in the particle basis for convenience. The covariant derivative 
takes the form
\begin{eqnarray}
 (D_\mu {B}_\nu)^{ijh}
   = \,
  \partial_\mu {B}_\nu^{ijh}
  + \Gamma^{i}_{\mu,l}\,{B}_\nu^{ljh}
  + \Gamma^{j}_{\mu,l}\,{B}_\nu^{ilh}
  + \Gamma^{h}_{\mu,l}\,{B}_\nu^{ijl}
  \,,
\end{eqnarray}
where again the chiral connection $\Gamma_\mu$ is needed. 
In order to keep track of the various flavor index contraction in the many terms of the chiral Lagrangian we use here a powerful notation already introduced by one of the authors in \cite{Lutz:2001yb}. The idea behind the notation is to introduce a few auxiliary objects in terms of which 
any interaction term can be written down in terms of simple $3 \times 3$ matrix products like it is the case in the baryon octet sector. Indeed this is achieved by the consideration of suitable 'dot' products of the decuplet fields. We need to discriminate the following three cases only 
\begin{eqnarray}
 (\bar{B}^{\mu} \cdot {B}_\nu)^i_j
  = \bar{B}^{\mu}_{jkl}\,{B}_\nu^{ikl}
  \,,\qquad
  (\bar{B}^{\mu} \cdot \Phi)^i_j = \epsilon^{kli}\,\bar{B}^{\mu}_{kmj}\,\Phi^m_l
  \,,\qquad
  (\Phi \cdot {B}_\nu)^i_j = \epsilon_{klj}\,\Phi^l_m\,{B}_\nu^{kmi}
  \,,
\end{eqnarray}
where any of such product yields a two-index object that transforms as a flavor octet field again. Note that it takes a bit of group theory that indeed all our terms in the chiral Lagrangian can be written down in such a notation. Given this fact, it is however, rather convenient to apply such a notation, since the painful write-down of flavor redundant terms can be avoided to a large extent. In \cite{Lutz:2001yb,LutzSemke2010} all terms at order $Q^2$ that are relevant for meson-baryon scattering were written down for the 
first time. Such terms were recently rediscovered in \cite{Jiang:2018mzd,Holmberg:2018dtv} using  a less transparent notation. The first partial list of $Q^2$ terms involving the baryon decuplet field was published in \cite{Tiburzi:2004rh}. 

We now turn to the symmetry preserving $Q^3$ terms that involve a decuplet  field. A 
complete list of 14 = 8+6  terms is readily worked out with
\allowdisplaybreaks[1]
\begin{eqnarray}
  && {\mathcal L}^{(3)}_{[10]\,[10]} = \tfrac{1}{2}\,v^{}_{1}\,\big(
  \tr\,(\bar{B}_{\tau}\cdot\,{U}^\nu)\,\gamma^\mu\,({H}_{\mu\nu}\cdot\,{B}^{\tau})
  + \text{h.c.}
  \big)
  \nonumber\\ &&  \quad \,
  + \tfrac{1}{2}\,v^{}_{2}\,\big(
  \tr\,(\bar{B}_{\lambda}\cdot\,{U}^{\lambda})\,\gamma^\mu\,({H}_{\mu\nu}\cdot\,{B}^{\nu})
  +\text{h.c.}
  \big)
  + \,\tfrac{1}{2}\,v^{}_{3}\,\big(
  \tr\,(\bar{B}^{\nu}\cdot\,{U}^{\lambda})\,\gamma^\mu\,({H}_{\mu\nu}\cdot\,{B}_{\lambda})
  + \text{h.c.}
  \big)
  \nonumber\\ &&  \quad \,
  + \,\tfrac{1}{4}\,v^{}_{4}\,\big(
  \tr\,(\bar{B}_{\tau}\cdot\,{U}_\lambda)\,\gamma^\lambda\,({H}_{\mu\nu}\cdot\,({D}^{\mu\nu}{B}^{\tau}))
  + \tr\,(({D}^{\mu\nu}\bar{B}_{\tau})\cdot\,{U}_\lambda)\,\gamma^\lambda\,({H}_{\mu\nu}\cdot\,{B}^{\tau})
  + \text{h.c.}
  \big)    
  \nonumber\\  && \quad \,
  + \,\tfrac{1}{2}\,v^{}_{5}\,\big(
  \tr\,(\bar{B}_{\tau}\cdot\,\sigma^{\lambda \mu}\,({D}^{\nu}{B}^{\tau}))\,[{U}_\lambda,\,{H}_{\mu \nu }]_+
  - \tr\,(({D}^{\nu}\bar{B}_{\tau})\cdot\,\sigma^{\lambda \mu}\,{B}^{\tau})\,[{U}_\lambda,\,{H}_{\mu \nu }]_+
  \big)
  \nonumber\\ &&  \quad \,
  +\, \tfrac{1}{4}\,v^{}_{6}\,\big(
  \tr\,(\bar{B}_{\tau}\cdot\,{U}_\lambda)\,\sigma^{\lambda\mu}\,({H}_{\mu\nu}\cdot\,({D}^{\nu}{B}^{\tau}))
  - \tr\,(({D}^{\nu}\bar{B}_{\tau})\cdot\,{U}_\lambda)\,\sigma^{\lambda\mu}\,({H}_{\mu\nu}\cdot\,{B}^{\tau})
  + \text{h.c.} \big)
  \nonumber\\ &&  \quad \,
  + \,\tfrac{1}{2}\,v^{}_{7}\,\big(
  \tr\,(\bar{B}_{\tau}\cdot\,\sigma^{\lambda \mu}\,({D}^{\nu}{B}^{\tau}))\,\tr\,{U}_\lambda\,{H}_{\mu \nu}
  - \tr\,(({D}^{\nu}\bar{B}_{\tau})\cdot\,\sigma^{\lambda \mu}\,{B}^{\tau})\,\tr\,{U}_\lambda\,{H}_{\mu \nu}
  \big)      
  \nonumber\\ &&  \quad \,
  +\, \tfrac{1}{4}\,v^{}_{8}\,\big(
  \tr\,(\bar{B}^{\mu}\cdot\,{U}_\lambda)\,({H}_{\mu\nu}\cdot\,({D}^{\lambda}{B}^{\nu}))
  - \tr\,(({D}^{\lambda}\bar{B}^{\mu})\cdot\,{U}_\lambda)\,({H}_{\mu\nu}\cdot\,{B}^{\nu})
  + \text{h.c.} \big)
  \,,
\end{eqnarray}
and
\begin{eqnarray}
  && {\mathcal L}^{(3)}_{[8]\,[10]} =  
  \tfrac{1}{2}\,w^{}_{1}\,\big(
  \tr\,\big( \bar{B}^\nu \cdot [{U}_\lambda,\,{H}_{\mu\nu}]_+ \big)\,i\,\sigma^{\lambda\mu}\gamma_5\,{B}
  +   \text{h.c.}
  \big)
  \nonumber\\ && \,
  + \tfrac{1}{4}\,w^{}_{2}\,\big(
  \tr\,\big( \bar{B}^\lambda \cdot[{U}_\lambda,\,{H}_{\mu\nu}]_+ \big)\,i\,\gamma^\mu\gamma_5\,({D}^\nu{B})
  - \tr\,\big( ({D}^\nu\bar{B}^\lambda) \cdot[{U}_\lambda,\,{H}_{\mu\nu}]_+ \big)\,i\,\gamma^\mu\gamma_5\,{B}
  +   \text{h.c.}
  \big)
  \nonumber\\ && \,
  + \tfrac{1}{4}\,w^{}_{3}\,\big(
  \tr\,\big( \bar{B}^\lambda \cdot[{U}_\mu,\,{H}_{\lambda\nu}]_+ \big)\,i\,\gamma^\mu\gamma_5\,({D}^\nu{B})
  - \tr\,\big( ({D}^\nu\bar{B}^\lambda) \cdot[{U}_\mu,\,{H}_{\lambda\nu}]_+ \big)\,i\,\gamma^\mu\gamma_5\,{B}
  +   \text{h.c.}
  \big)
  \nonumber\\
  && \, +\, 
  \tfrac{1}{2}\,w^{}_{4}\,\big(
  \tr\,\big( \bar{B}^\nu \cdot [{U}_\lambda,\,{H}_{\mu\nu}]_- \big)\,i\,\sigma^{\lambda\mu}\gamma_5\,{B}
  +   \text{h.c.}
  \big)
  \nonumber\\ && \,
  + \tfrac{1}{4}\,w^{}_{5}\,\big(
  \tr\,\big( \bar{B}^\lambda \cdot[{U}_\lambda,\,{H}_{\mu\nu}]_- \big)\,i\,\gamma^\mu\gamma_5\,({D}^\nu{B})
  - \tr\,\big( ({D}^\nu\bar{B}^\lambda) \cdot[{U}_\lambda,\,{H}_{\mu\nu}]_- \big)\,i\,\gamma^\mu\gamma_5\,{B}
  +   \text{h.c.}
  \big)
  \nonumber\\ && \,
  + \tfrac{1}{4}\,w^{}_{6}\,\big(
  \tr\,\big( \bar{B}^\lambda \cdot[{U}_\mu,\,{H}_{\lambda\nu}]_- \big)\,i\,\gamma^\mu\gamma_5\,({D}^\nu{B})
  - \tr\,\big( ({D}^\nu\bar{B}^\lambda) \cdot[{U}_\mu,\,{H}_{\lambda\nu}]_- \big)\,i\,\gamma^\mu\gamma_5\,{B}
  +   \text{h.c.}
  \big)  \,.
\end{eqnarray}
We observe a significant mismatch with the number of seven terms claimed in \cite{Jiang:2018mzd}. 

\section{Correlation function from the chiral Lagrangian}

We consider QCD's axial-vector and vector currents,
\begin{eqnarray}
&& A_\mu^{(a)}(x) = \bar \Psi (x)\,\gamma_\mu \,\gamma_5\,\frac{\lambda_a}{2}\,\Psi(x) \,, \qquad \qquad
V^{(a)}_\mu(x) = \bar \Psi (x)\,\gamma_\mu\,\frac{\lambda_a}{2}\,\Psi(x) \,,
\label{def-amu}
\end{eqnarray}
where we recall their definitions in terms of the
Heisenberg quark-field operators $\Psi(x)$. With $\lambda_a$ we
denote the Gell-Mann flavor matrices. Our target is an evaluation of the following matrix elements
\begin{eqnarray}
  C^{(abe)}_{\mu\nu\lambda}(q,q') 
  = \int d^4{x}\,d^4{y}\,e^{+i\,q\cdot({x}-{y})}   \,e^{+i\,q'\cdot y}\,
  \bra{ \,\bar p,\,\bar \chi} \,
  {\cal T}\,A_\mu^{(a)}(x)\,A_\nu^{(b)}(0)\,V_\lambda^{(e)}(y) \,\ket{p,\,\chi} \,,
  \label{def-C}
 \end{eqnarray}
in the baryon ground states. Here the spin projections of the initial and final baryon states we denote by $\chi$ and $\bar \chi$. Similarly the initial and final three momenta of the states are $p$ and $\bar p$. The flavor structure in (\ref{def-C}) is incomplete since also the initial and final baryon states come in different flavor copies. We return to this issue below in more detail. Given the chiral Lagrangian, it is well defined how to derive the contributions to such matrix elements in application  of the  classical matrices of source functions, $a_\mu$ and $v_\mu$. 

The particular correlation function is chosen as to selectively probe our $Q^3$ terms. This is so since any such term in the chiral Lagrangian is linear in the $U_\lambda$ field but also in the $H_{\mu \nu}$ field. Upon an expansion of those building blocks in powers of the meson fields one finds 
\begin{eqnarray}
i\,U_\mu = a_\mu + \cdots \,,\qquad \qquad \qquad
i\,H_{\mu \nu} = \big[ v_\mu, \, a_\nu \big]_- + \big[ v_\nu, \, a_\mu \big]_- + \cdots \,.
 \end{eqnarray}
From here we conclude that the tree-level evaluation of the chiral Lagrangian 
is charcterized by the symmetry conserving $Q^3$ terms, as anticipated above. 

The motivation for our study of this correlation function is twofold. First, it serves as a convenient tool as to verify whether we use only independent sets of the symmetry conserving $Q^3$ terms. We checked for the flavor octet case, that any additional term leads to a contribution that can be linear combined in terms of the 10 terms originally used in \cite{Lutz2002a} and confirmed later in \cite{Frink:2006hx,Oller:2007qd}. An analogous 
computation consolidates our claim about the smallest set of independent terms in the 
decuplet sector. Second, such a correlation function can be scrutinized also in large-$N_c$ QCD. This will lead to sum rules amongst the set of low-energy constants introduced in this work. We will turn to this issue in the next section. 

We close this section with explicit results for the correlation function. It suffices to evaluate the matrix elements in the strict flavor SU(3) limit. In this case a baryon octet or a decuplet state
\begin{eqnarray}
\ket{p, \,\chi,\, c }\,, \qquad \qquad \qquad \ket{p, \,\chi, \,klm } \,,
\label{def-states}
\end{eqnarray}
is specified by its three-momentum $p$ and the flavor indices $c=1,\cdots ,8$ or $k,l,m=1,2,3$. The spin-polarization label is $\chi = 1,2$ for the octet and $\chi =1,\cdots ,4$ for the decuplet states. In order to discriminate flavor structures from the currents versus those from the baryon states
we introduce the operator
\begin{eqnarray}
  {\mathcal O}^{(abe)}_{ijh}(q,q') 
  = \int d^4{x}\,d^4{y}\,e^{+i\,q\cdot({x}-{y})}   \,e^{+i\,q'\cdot y}\,
  {\cal T}\,A_i^{(a)}(x)\,A_j^{(b)}(0)\,V_h^{(e)}(y)  \,,
  \label{def-O}
 \end{eqnarray}
which matrix elements in the baryon states (\ref{def-states}) are considered in the following. Note that in (\ref{def-O}) we already focus on the space components of the three currents. From the study of such components the anticipated large-$N_c$ sum rules for the low-energy constants, the main target of our work, can be derived. 

Since we will encounter many flavor indices in our work, which either run from one to three or from one to eight, we found it useful to split the alphabet into two parts. 
We use the Roman small letters from $a$ to $g$ for flavor indices with $a=1,..,8$ and letters from $h$ to $z$ for indices with $h=1,..,3$. With this convention it is easily confirmed over which range a given flavor index goes.

We are now prepared to present results for the matrix elements introduced with (\ref{def-C}).
A somewhat tedious but straightforward evaluation leads to the explicit results
\allowdisplaybreaks[1]
\begin{eqnarray}
  && \qquad \qquad \qquad \qquad \qquad \qquad
  \bra{ \bar{p}, \bar\chi, d}\, {\mathcal O}^{(abe)}_{ijh}(q,q') \,\ket{ p, \chi, c}
  \nonumber\\ \nonumber\\
  && = \,
  \bar{u}(\bar{p},\bar\chi)\frac{
    2\,g^{{i}{j}}\,\gamma^{h}
    + g^{{i}{h}}\,\gamma^{j} + g^{{j}{h}}\,\gamma^{i}
  }{4}\,{u}({p},\chi)
  \,\Big\{
      - (u^{}_{1}+u^{}_{2})\,\delta_{ab}\,d_{{d}{c}e}
      - 3\,(u^{}_{1}+u^{}_{2})\,d_{abg}\,d_{efg}\,d_{{d}{c}f}
\nonumber\\
&& \qquad -\, (u^{}_{1}-u^{}_{2})\,\delta_{ab}\,if_{{d}{c}e}
      - 3\,(u^{}_{1}-u^{}_{2})\,d_{abg}\,d_{efg}\,if_{{d}{c}f}
       + (u^{}_{1}+u^{}_{2})\,(\delta_{ae}\,d_{b{d}{c}} + \delta_{be}\,d_{a{d}{c}})
\nonumber\\
&& \qquad
      + \,(u^{}_{1}-u^{}_{2})\,(\delta_{ae}\,if_{b{d}{c}} + \delta_{be}\,if_{a{d}{c}})
      - \tfrac{1}{2}\,u^{}_{3}\,(
      \delta_{a{d}}\,if_{be{c}} + \delta_{b{d}}\,if_{ae{c}}
      - \delta_{a{c}}\,if_{be{d}} - \delta_{b{c}}\,if_{ae{d}}
      )
    \Big\}
\nonumber\\ && 
  +\, \bar{u}(\bar{p},\bar\chi)\,\frac{
    g^{{i}{h}}\,\gamma^{j} - g^{{j}{h}}\,\gamma^{i}
  }{4}\,{u}({p},\chi)
  \,\Big\{
       (u^{}_{1}+u^{}_{2})\,f_{abg}\,f_{efg}\,d_{{d}{c}f}
           + (u^{}_{1}-u^{}_{2})\,f_{abg}\,f_{efg}\,if_{{d}{c}f}
\nonumber\\
&& \qquad
      -\, \tfrac{1}{2}\,u^{}_{3}\,(
      \delta_{a{d}}\,if_{be{c}} - \delta_{b{d}}\,if_{ae{c}}
      - \delta_{a{c}}\,if_{be{d}} + \delta_{b{c}}\,if_{ae{d}}
      )
    \Big\}
\nonumber\\ && 
  +\, \bar{u}(\bar{p},\bar\chi)\,\frac{
    i\,\sigma^{{i}{h}}\,\delta_{jk}
    + i\,\sigma^{{j}{h}}\,\delta_{ik}
  }{8}\,{u}({p},\chi)
  \,(\bar{p}+{p})^{k}
  \,\Big\{
      - (u^{}_{7} + u^{}_{8} - u^{}_{9})\,d_{abg}\,i\,f_{efg}\,d_{{d}{c}f}
\nonumber\\
&& \qquad
      - \,(u^{}_{7} - u^{}_{8})\,d_{abg}\,i\,f_{efg}\,i\,f_{{d}{c}f}
      - \tfrac{1}{2}\,u^{}_{9}\,(
          \delta_{a{d}}\,i\,f_{be{c}} + \delta_{b{d}}\,i\,f_{ae{c}}
            + \delta_{a{c}}\,i\,f_{be{d}} + \delta_{b{c}}\,i\,f_{ae{d}} )
\Big\}
\nonumber\\ && 
  +\, \bar{u}(\bar{p},\bar\chi)\, \frac{2\, i\,\sigma^{{i}{j}}\,\delta_{hk}
       + i\,\sigma^{{i}{h}}\,\delta_{jk}
        - i\,\sigma^{{j}{h}}\,\delta_{ik}
      }{8}
   \,{u}({p},\chi)
  \,\Big\{    - (
      \tfrac{4}{3}\,u^{}_{7} + \tfrac{4}{3}\,u^{}_{8}
      - \tfrac{1}{3}\,u^{}_{9}
      + 2\,u^{}_{10}
      )\,if_{abe}\,\delta_{{d}c}
\nonumber \\
&& \qquad 
      +\, (u^{}_{7} + u^{}_{8}  - \,u^{}_{9})\,(if_{aeg}\,d_{bgf} - if_{beg}\,d_{agf})\,d_{{d}{c}f}
\nonumber \\
&& \qquad      +\, (u^{}_{7} - u^{}_{8})\,(if_{aeg}\,d_{bgf} - if_{beg}\,d_{agf})\,i\,f_{{d}{c}f}
\nonumber \\
&& \qquad
      -\, \tfrac{1}{2}\,u^{}_{9}\,(
          \delta_{a{d}}\,if_{be{c}} -\delta_{b{d}}\,if_{ae{c}}
            + \delta_{a{c}}\,if_{be{d}} - \delta_{b{c}}\,if_{ae{d}}
        )
    \Big\} \,(\bar{p}+{p})^{k}
\nonumber\\ && 
  +\, \bar{u}(\bar{p},\bar\chi)\,\frac{
    \bar{p}^{i}\,\gamma^{j}
    + \bar{p}^{j}\,\gamma^{i}
  }{2}\,{u}({p},\chi)\,(\bar{p}+p)^{h}
  \,\Big\{
      - (u^{}_{4} + u^{}_{5})\,\delta_{ab}\,d_{{d}{c}e}
           - 3\,(u^{}_{4} + u^{}_{5})\,d_{abg}\,d_{efg}\,d_{{d}{c}f}
\nonumber\\
&& \qquad
      -\, (u^{}_{4} - u^{}_{5})\,\delta_{ab}\,if_{{d}{c}e}
      - 3\,(u^{}_{4} - u^{}_{5})\,d_{abg}\,d_{efg}\,if_{{d}{c}f}
     +  (u^{}_{4} + u^{}_{5})\,(\delta_{ae}\,d_{b{d}{c}} + \delta_{be}\,d_{a{d}{c}})
\nonumber\\
&& \qquad
      +\,  (u^{}_{4} - u^{}_{5})\,(\delta_{ae}\,i\,f_{b{d}{c}} + \delta_{be}\,if_{a{d}{c}})
\nonumber\\
&& \qquad
      -\, \tfrac{1}{2}\,u^{}_{6}\,\big(
      \delta_{a{d}}\,if_{be{c}} + \delta_{b{d}}\,if_{ae{c}}
      - \delta_{a{c}}\,if_{be{d}} - \delta_{b{c}}\,\,if_{ae{d}}
      \big)
    \Big\}
   \nonumber\\ && 
   - \,
     \bar{u}(\bar{p},\bar\chi)\,\frac{
       \bar{p}^{i}\,\gamma^{j}
       - \bar{p}^{j}\,\gamma^{i}
     }{2}\,{u}({p},\chi)
   \,\Big\{
       (u^{}_{4} + u^{}_{5})\,f_{abg}\,f_{efg}\,d_{{d}{c}f}
       + (u^{}_{4} - u^{}_{5})\,f_{abg}\,f_{efg}\,if_{{d}{c}f}
 \nonumber\\
 && \qquad
       - \,\tfrac{1}{2}\,u^{}_{6}\,(
       \delta_{a{d}}\,if_{be{c}} - \delta_{b{d}}\,if_{ae{c}}
       - \delta_{a{c}}\,if_{be{d}} + \delta_{b{c}}\,if_{ae{d}}
       )
     \Big\}\,(\bar{p}-p)^{h}
  \,.
  \label{res-15}
\end{eqnarray}
Corresponding expressions for  matrix elements in the baryon decuplet states are collected in Appendix A. We wish to emphasize that the computation of such matrix elements serves as a powerful consistency check whether the terms of the chiral Lagrangian were constructed properly. Our results (\ref{res-15}) show that all terms shown are independent, i.e. it is not possible to eliminate any term.

\newpage

\section{Current correlation function in large-$N_c$ QCD}

Consider ${\mathcal O}_{QCD}$ to be the time ordered product of any combination of local currents in large-$N_c$ QCD, where (\ref{def-O}) may serve as a specific example for $N_c = 3$. 
The generic form of the large-$N_c$ operator expansion can be taken as
\begin{eqnarray}
  \bra{\,\bar p,\bar\chi\,}\,{\mathcal O}_{QCD}\,
\ket{p,\,\chi\,} = \sum_{n=0}^\infty\, c_n( \bar p, p)\,
\roundbra{\bar\chi \,} \,{\mathcal O}^{(n)}_{\rm static} \,
\roundket{\chi\,} \,,
\label{def-largeN-expansion}
\end{eqnarray}
where it is important to note that unlike the physical baryon states, $\ket{p,\,\chi\,}$, 
the effective baryon states, $\roundket{\chi\,}$,  do not depend on the three-momentum $p$. All dynamical information in (\ref{def-largeN-expansion}) is moved into appropriate coefficient functions $c_n(\bar p, p)$. Moreover, in the decomposition (\ref{def-largeN-expansion}) the coefficients $ c_n(\bar p, p) $ depend on neither the flavor 
nor the spin quantum number of the initial or the final baryon state. The merit of (\ref{def-largeN-expansion}) lies in the fact that the contributions on its right-hand-side can be sorted according to their relevance at large values of $N_c$. 

The effective baryon states $\roundket{c, \chi}$  and $\roundket{klm, \chi}$  have a mean-field structure that can be generated in terms of effective quark operators. They correspond to 
the baryon states already introduced with  (\ref{def-states}) for the particular choice $N_c =3$. A complete set of color-neutral one-body operators may be constructed in terms of the very same static quark operators
\begin{eqnarray}
&& \quarknumberoperator = q^\dagger ( \one  \otimes \one  \otimes \one )\,q \,, \qquad  \qquad \;\;\,
J_i = q^\dagger \Big(\frac{\sigma_i }{2} \otimes \one \otimes \one \Big)\, q \,,
\nonumber\\
&& T^a = q^\dagger \Big(\one \otimes \frac{\lambda_a}{2} \otimes \one \Big)\, q\, ,\qquad \quad \;\;
G^a_i = q^\dagger \Big( \frac{\sigma_i}{2} \otimes \frac{\lambda_a}{2} \otimes \one \Big)\, q\,,
\label{def:one-body-operators}
\end{eqnarray}
with operators $q=(u,d,s)^T$ introduced for the up, down and strange quarks. With $\lambda_a$ we
denote the Gell-Mann matrices. While the action of any of the spin-flavor operators introduced in (\ref{def:one-body-operators}) on the tower of large-$N_c$ states is quite involved at large $N_c \neq 3$ matters turn quite simple and straightforward at the physical value $N_c = 3$. For this physical case where there is a flavor octet with spin-one-half or a flavor decuplet with spin-three-half only, we  recall the well established  results of \cite{Lutz2002a,LutzSemke2010} with
\begin{eqnarray}
&& \quarknumberoperator \,\roundket{c,\chi} = 3\,\roundket{c,\chi}\,,
\qquad \qquad 
\nonumber \\
&& J_i \,\roundket{c,\chi}=\frac{1}{2}\, \sigma^{(i)}_{{\bar \chi} \chi}\, \roundket{c,{\bar \chi}}\,,
\qquad \qquad 
T^a\, \roundket{c,\chi} = i\,f_{cda}\, \roundket{d,\chi}\,,
\nonumber \\
&& G^{a}_i\, \roundket{c,\chi} =  \sigma^{(i)}_{{\bar \chi} \chi}\, \Big(\frac12\,d_{cda} + \frac{i}{3}\, f_{cda}\Big)\,
\roundket{d,{\bar \chi}} + \frac{1}{2\sqrt{2}}\, S^{(i)}_{{\bar \chi} \chi}\, \Lambda_{ac}^{klm}
\, \roundket{klm,{\bar \chi}}\,,
\nonumber \\
\nonumber \\
&& \quarknumberoperator \,\roundket{klm, \chi}=3\,\roundket{klm, \chi}\,,
\qquad \qquad 
\nonumber\\
&& J_i \,\roundket{klm, \chi}= \frac{3}{2}\,\Big(\vec S \,\sigma_i\, \vec S^\dagger \Big)_{{\bar \chi} \chi}\, \roundket{klm, {\bar \chi}},
\qquad \qquad
 T^a \,\roundket{klm, \chi}= \frac{3}{2}\,\Lambda^{a,nop}_{klm}\, \roundket{nop,\chi},
\nonumber \\
&& G^a_i \,\roundket{klm, \chi}= \frac34 \, \Big(\vec S\,\sigma_i\, \vec S^\dagger \Big)_{{\bar \chi} \chi}\,
\Lambda^{a,nop}_{klm}\, \roundket{nop, {\bar \chi}}  +
\frac{1}{2\sqrt{2}} \, \Big(S^{\dagger}_i \,\Big)_{{\bar \chi} \chi}\, \Lambda^{ac}_{klm}\, \roundket{c,{\bar \chi}}\,,
\label{result:one-body-operators}
\end{eqnarray}
with the Pauli matrices $\sigma_i$ and the spin-transition matrices $S_i$ characterized by
\begin{eqnarray}
&& S^\dagger_i\, S_j= \delta_{ij} - \frac{1}{3}\sigma_i \sigma_j \,, \qquad S_i\,\sigma_j - S_j\,\sigma_i = -i\,\varepsilon_{ijk} \,S_k\,,
\qquad \vec S\cdot  \vec S^\dagger= \one_{(4\times 4)}\,,
\nonumber\\
&& \vec S^\dagger \cdot  \vec S =2\, \one_{(2\times 2)}\,, \qquad \vec S \cdot \vec \sigma = 0 \,,\qquad
\epsilon_{ijk}\,S_i\,S^\dagger_j = i\,\vec S \,\sigma_k\,\vec S^\dagger\,.
\label{def:spin-transition-matrices}
\end{eqnarray}
We recall some instrumental flavor structures 
\begin{eqnarray}
&& \Lambda_{ab}^{klm} = \Big[\varepsilon_{ijk}\, \lambda^{(a)}_{li}\,
\lambda^{(b)}_{mj} \,\Big]_{\mathrm{sym}(klm)}\,,\qquad \quad
\delta^{\,klm}_{\,nop} \;\;= \Big[\delta_{kn}\,\delta_{lo}\,\delta_{mp} \,\Big]_{\mathrm{sym}(nop)}\,,
\nonumber\\
&&  \Lambda^{ab}_{klm} = \Big[\varepsilon_{ijk}\, \lambda^{(a)}_{il}\,
\lambda^{(b)}_{jm} \,\Big]_{\mathrm{sym}(klm)}\,,\qquad \quad
\Lambda^{a,klm}_{nop} = \Big[\lambda^{(a)}_{kn} \delta_{lo} \,\delta_{mp}\Big]_{\mathrm{sym}(nop)}\,,
\label{def:flavor-transition-matrices}
\end{eqnarray}
that occur
frequently in our previous and current works \cite{Lutz:2010se,Lutz:2014jja,Lutz:2018cqo}. 

In the sum of (\ref{def-largeN-expansion}) there are infinitely many terms one may write down. The static operators ${\mathcal O}_{\rm static}^{(n)}$ 
are finite products of the one-body operators $J_i\,, T^a$ and $G^a_i$. 
In contrast the counting of $N_c$ factors is intricate since there is
a subtle balance of suppression and enhancement effects.
An $r$-body operator consisting of the $r$ products of any of the spin and flavor operators receives the
suppression factor $N_c^{-r}$. This is counteracted by enhancement factors for the
flavor and spin-flavor operators $T^a$ and $G^a_i$ that are
produced by taking baryon matrix elements at $N_c \neq 3$.
Altogether this leads to the effective scaling laws \cite{Dashen1994}
\begin{eqnarray}
J_i \sim \frac{1}{N_c} \,, \qquad \quad T^a \sim N^0_c \,, \qquad \quad G^a_i \sim N^0_c \,.
\label{effective-counting}
\end{eqnarray}
According to (\ref{effective-counting}) there are an infinite number of terms contributing at a given order in the
the $1/N_c$ expansion. Taking higher products of flavor and spin-flavor operators does not reduce the $N_c$
scaling power. A systematic $1/N_c$ expansion is made possible by a set of operator
identities \cite{Dashen1994,Lutz:2010se}, that allows a systematic summation of the infinite number
of relevant terms.  As a 
consequence of the SU(6) Lie algebra any commutator of one-body operators can be  
expressed in terms of one-body operators again. Therefore it suffices to consider anti commutators of the one-body operators \cite{Dashen1994,Lutz:2010se}. For instance, consider the following two  
identities that hold in matrix elements of the baryon states 
\begin{eqnarray}
  d_{gab}\,[T_a,\,T_b]_+
  && \, = \,
  - 2\,T_g
    + 2\,[J^i,\,G^i_g]_+
  \,,\nonumber\\
  d_{gab}\,[G^i_a,\,G^j_b]_+
  && \, = \,
  \tfrac{1}{3}\,\delta^{ij}\,\big(
    \tfrac{9}{2}
  \,T_g - \tfrac{3}{2}\,[J^k,\,G^k_g]_+\big)
  + \tfrac{1}{6}\,\big([J^i,\,G^j_g]_++[J^j,\,G^i_g]_+\big)\,.
\end{eqnarray}
Altogether the expansion scheme  is implied by two reduction rules:
\begin{itemize}
\item All operator products in which two flavor indices are contracted using $\delta_{ab}$,
$f_{abc}$ or $d_{abc}$ or two spin indices on $G$'s are contracted using $\delta_{ij}$ or $\varepsilon_{ijk}$
can be eliminated.
\item All operator products in which two flavor indices are contracted using symmetric or antisymmetric
combinations of two different $d$ and/or $f$ symbols can be eliminated. The only exception to this rule is
the antisymmetric
combination $f_{acg}\,d_{bch}-f_{bcg}\,d_{ach}$.
\end{itemize}
As a consequence the infinite tower of spin-flavor operators truncates at any given order in the $1/N_c$ expansion.
We can now turn to the $1/N_c$ expansion of the baryon matrix elements of our specific product of QCD's axial-vector and vector currents.
In application of the operator reduction rules, the baryon matrix elements of time-ordered products of the current
operators are expanded in powers of the effective one-body operators according to the counting rule (\ref{effective-counting})
supplemented by the reduction rules.

\section{Sum rules for the low-energy constants}

As compared to previous works \cite{Luty1994,Dashen1995,Lutz:2010se,Lutz:2018cqo} that dealt with correlation functions of one or two currents only, it turned out that the systematic construction of the large-$N_c$ operator hierachy for 
the correlation function of three currents is considerably more involved. While it is straightforward to write down a set of operators to a given order almost any single term cannot be matched to the matrix elements as implied by the chiral Lagrangian. This is so since the role of charge conjugation and parity invariances is not so transparent 
in the given frame work. We derived our operators by considering all possible combinations 
and then performed the matching in application of a suitable computer algebra code. This then generated the following leading order decomposition 
\begin{eqnarray}
 && {\cal O}^{ijh}_{abe}
   = \delta^{(ij)_+}_{h}\,\Big\{
  \hat{g}_{1}\,\big(
  \delta_{ab}\,T_e
  - (\delta_{ae}\,T_b + \delta_{be}\,T_a)
  + 3\,d_{abg}\,d_{efg}\,T_f
  \big)
  \nonumber\\ && \, \qquad\qquad
  -\, \tfrac{1}{2}\,\hat{g}_{4}\,\big\{
  d_{aeg}\,[J^l,\,([T_g,\,G^l_b]_+ - [T_b,\,G^l_g]_+)]_+
  + d_{beg}\,[J^l,\,([T_g,\,G^l_a]_+ - [T_a,\,G^l_g]_+)]_+
  \nonumber\\ && \, \qquad\qquad\qquad
  - 2\,d_{abg}\,[J^l,\,([T_g,\,G^l_e]_+ - [T_e,\,G^l_g]_+)]_+
  \big\}  \Big\}
\nonumber\\   
&& \qquad +\,  (\bar{p}+{p})^q\,\Big\{\big(i\,\epsilon^{iju}\,\delta_{hq} + \delta^{(ij)_-}_{(vq)_-}\,i\,\epsilon^{huv}\big)\,\Big[
  \hat{g}_{2}\,(i\,f_{aeg}\,d_{gbf} - i\,f_{beg}\,d_{gaf})\,G^u_f
  + \hat{g}_{5}\,i\,f_{abe}\,J^u
    \nonumber\\ && \, \qquad\qquad \qquad
  + \,\hat{g}_{6}\,(i\,f_{aeg}\,d_{bfg} - i\,f_{beg}\,d_{afg})\,[J^u,\,T_f]_+
  \Big]
   \nonumber\\ && \, \qquad \qquad
  +\, \delta^{(ij)_+}_{(vq)_+}\,i\,\epsilon^{huv}\,\Big[
  \hat{g}_{3}\,d_{abg}\,i\,f_{feg}\,G^u_f
  +\hat{g}_{7}\,d_{abg}\,if_{efg}\,[J^u,\,T_f]_+
  \Big]
  \Big\}
   \,,
\label{QCD-identity-AVA}
\end{eqnarray}
where the parameters $\hat g_{1-3}$ and $\hat g_{4-7}$ are relevant at leading and subleading orders respectively. In (\ref{QCD-identity-AVA}) we use the notation
\begin{eqnarray}
\delta_{(mn)_\pm}^{(ij)_\pm} = 
  \tfrac{1}{2}\,(\delta_{mi}\,\delta_{nj} \pm \delta_{mj}\,\delta_{ni})\,,\quad
  \delta^{(ij)_\pm}_h =   \tfrac{1}{2\,(\bar{M}+{M})}\,\big(\bar p^i \,\bar p^h + p^i \, p^h\big) \,\big(\bar p+ p\big)^j \pm (i \leftrightarrow j) \,.
\end{eqnarray}

\begin{table}[t]
  \centering
  \begin{tabular}{l|l|l}
$ u^{}_{1} = 0                                                    $&$ v^{}_{1} = 0                                 $&$ w^{}_{1} = 0  $ \\
$ u^{}_{2} = 0                                                      $&$ v^{}_{2} = 0                                 $&$ w^{}_{2} = - 4\,\hat{g}_{2}$ \\
$ u^{}_{3} = 0                                                      $&$ v^{}_{3} = 0                                 $&$ w^{}_{3} =  4\,\hat{g}_{2} $\\
$ u^{}_{4} = \tfrac{1}{2}\,\hat{g}_{1} + \tfrac{1}{2}\,\hat{g}_{4}  $&$ v^{}_{4} = - 3\,\hat{g}_{1}                  $&$ w^{}_{4} = 0 $\\
$ u^{}_{5} = -\tfrac{1}{2}\,\hat{g}_{1} - \tfrac{1}{2}\,\hat{g}_{4} $&$ v^{}_{5} = -3\,\hat{g}_{2} - 18\,\hat{g}_{6} $&$ w^{}_{5} = 0 $\\
$ u^{}_{6} = 3\,\hat{g}_{4}                                         $&$ v^{}_{6} = 0                                 $&$ w^{}_{6} = 0 $\\ 
$ u^{}_{7} = \tfrac{1}{3}\,\hat{g}_{2} - 2\,\hat{g}_{6}             $&$ v^{}_{7} = 2\,\hat{g}_{2} + 3\,\hat{g}_{6} + 12\,\hat{g}_{6} \qquad \qquad$& \\
$ u^{}_{8} = \tfrac{5}{3}\,\hat{g}_{2} + 2\,\hat{g}_{6}             $&$ v^{}_{8} = 0$& \\
$ u^{}_{9} = 0                                                      $& & \\
$ u^{}_{10}= -\tfrac{4}{3}\,\hat{g}_{2} - \hat{g}_{5}     \qquad \qquad          $& & \\
     \end{tabular}
  \caption{Matching of the large-$N_c$ operators to the LEC. }
  \label{tab:algebraic-sol}
\end{table}
Owing to the matching condition
\begin{eqnarray}
  \hat g_2 + \hat g_3 = 0 \,,
  \label{matching-condition-1}
\end{eqnarray}
there are two leading order operators only.  This is a non-trivial result in view of the fact that one may write down many more leading order operators. For instance consider the particular term 
\begin{eqnarray}
  \delta^{(ij)_+}_{h}\,\Big(  2\,[[T_a,\,T_b]_+,\,T_e]_+
  - ([[T_a,\,T_e]_+,\,T_b]_+ + [[T_b,\,T_e]_+,\,T_a]_+) \Big) \,,
  \label{eliminate}
\end{eqnarray}
which matrix elements can be shown to be proportional to the matrix elements of the  operator associated with $\hat g_1$. At subleading order we find three additional operators only. Here the matching condition
\begin{eqnarray}
  \hat g_6 + \hat g_7 = 0 \,,
\end{eqnarray}
eliminates one term.

The number of independent coupling constants in the chiral Lagrangian is 24. At leading order in the $1/N_c$ expansion all of them can be expressed in terms of $\hat g_1$ and $\hat g_2$ as detailed in Tab. \ref{tab:algebraic-sol}. At subleading order the additional three parameters $\hat g_4, \hat g_5 $ and $\hat g_6$ enter. The desired sum rules follow upon eliminating the parameters $\hat g_n$. There are 15 common sum rules applicable at LO and NLO
\begin{eqnarray}
&&  u^{}_{1,2,3} = 0 = u^{}_{9}   \,,\qquad \quad 
    v^{}_{1,2,3} = 0 = v^{}_{6,8}\,, \qquad \quad w^{}_{4,5,6} = 0=  w^{}_1\,,
\nonumber\\
&& u^{}_{5} =  -u^{}_{4}\,,\qquad \qquad \quad \;
  w^{}_{3}
  = -w^{}_{2}
  \,.
\end{eqnarray}
They are supplemented by 
4 and 7 additional sum rules at NLO and LO respectively as
\begin{eqnarray}
&&   v^{}_{5} = 6\,u^{}_{7}-3\,u^{}_{8}
  \,,\qquad \qquad \qquad 
  v^{}_{4} = 6\,u^{}_{5}+u^{}_{6}
  \,,
\nonumber\\ 
&&   v^{}_{7} = -6\,u^{}_{7}-3\,u^{}_{10}
  \,,\qquad \qquad \quad \!
  w^{}_{3}
  = 2\,(u^{}_{7}+u^{}_{8}) \,,
\end{eqnarray}
and
\begin{eqnarray}
&&   v^{}_{5}
  = -9\,u^{}_{7}
  = \tfrac{9}{4}\,u^{}_{10}
  \,,\qquad \qquad 
  v^{}_{4} = 6\,u^{}_{5}
  \,,\qquad \qquad
  u^{}_{6}=0
  \,,
\nonumber\\ && 
  v^{}_{7}
  = 6\,u^{}_{7}
  \,,\qquad \qquad \qquad 
  w^{}_{3}
  = 12\,u^{}_{7}
  = \tfrac{12}{5}\,u^{}_{8}
  \,.
\end{eqnarray}

\newpage

\section{Summary}

In this work we further prepared the ground for realistic applications of the chiral Lagrangian with the baryon octet and the baryon decuplet fields. For the first time all symmetry preserving $Q^3$ counter terms were constructed as they are relevant for any two-body meson-baryon reaction process. Altogether we find 24 terms. In order to pave the way towards applications of this set of low-energy parameters we derived a set of sum rules. We considered matrix elements of a correlation function with two axial-vector and one vector currents in the baryon ground states as they arise in QCD at a large number of colors ($N_c$). From a systematic operator expansion thereof we deduced our set of 22 sum rules valid at leading order in the $1/N_c$ expansion. At subleading order there remain 19 relations.
With our result we now deem it feasible to perform significant coupled-channel studies of meson-baryon scattering processes  considering channels with the baryon octet and decuplet fields on an equal footing as it is requested by large-$N_c$ QCD.

\vskip0.3cm
{\bfseries{Acknowledgments}}
\vskip0.3cm
Y. Heo and C. Kobdaj acknowledge partial support from Suranaree University of Technology, the Office of the Higher Education Commission under NRU project 
of Thailand (SUT-COE: High Energy Physics and Astrophysics) and  SUT-CHE-NRU (Grant No. FtR.11/2561).

\newpage

\section{Appendix A}

\begin{eqnarray}
  &&  \qquad \qquad \qquad \qquad \qquad \qquad
  \bra{\bar{p},\,\bar \chi,\,nop\, }\, {\mathcal O}^{(abe)}_{ijh}(q,q') \,\ket{p, \,\chi,\, klm\, }
  \nonumber\\ \nonumber\\
  && = \,
  -\,\tfrac{1}{4}\,
  \delta^{{nop}}_{xyz}\,\Lambda^{d,xyz}_{klm}\,
  \big( 3\,d_{abg}\,d_{edg}
      + \delta_{ab}\,\delta_{ed}
      - \delta_{ae}\,\delta_{bd} - \delta_{ad}\,\delta_{be} \big)
  \,\Big\{
  \nonumber\\ 
  && \qquad 
      +\, \tfrac{1}{2}\,v^{}_{1}  \bar{u}_{\tau}(\bar{p},\bar\chi)\,\big(
      2\,\gamma^{{h}}\,g^{{i}{j}}
      + \gamma^{{i}}\,g^{{h}{j}}
      + \gamma^{{j}}\,g^{{h}{i}}
      \big)\,{u}^\tau({p},\chi)\,
\nonumber\\
&& \qquad      + \, \tfrac{1}{2}\,v^{}_{2}\left(
          \bar{u}^{{i}}(\bar{p},\bar\chi)\,\gamma^{{j}}\,{u}^{h}({p},\chi)
          + \bar{u}^{{j}}(\bar{p},\bar\chi)\,\gamma^{{i}}\,{u}^{h}({p},\chi)
          + \bar{u}^{{i}}(\bar{p},\bar\chi)\,\gamma^{{h}}\,{u}^{j}({p},\chi)
          + \bar{u}^{{j}}(\bar{p},\bar\chi)\,\gamma^{{h}}\,{u}^{i}({p},\chi)
        \right)\,
\nonumber\\
&& \qquad 
      + \, \tfrac{1}{2}\,v^{}_{3}\left(
          \bar{u}^{{i}}(\bar{p},\bar\chi)\,\gamma^{{h}}\,{u}^{j}({p},\chi)
          + \bar{u}^{{j}}(\bar{p},\bar\chi)\,\gamma^{{h}}\,{u}^{i}({p},\chi)
          + \bar{u}^{{h}}(\bar{p},\bar\chi)\,\gamma^{{i}}\,{u}^{j}({p},\chi)
          + \bar{u}^{{h}}(\bar{p},\bar\chi)\,\gamma^{{j}}\,{u}^{i}({p},\chi)
        \right)\,
    \Big\}\,
  \nonumber\\ && \,
  -\,\tfrac{1}{4}\,
  \delta^{{nop}}_{xyz}\,\Lambda^{d,xyz}_{klm}\, f_{abg}\,f_{edg}
  \,\Big\{
       \tfrac{1}{2}\,v^{}_{1}\,\bar{u}_{\tau}(\bar{p},\bar\chi)\,\big(
      \gamma^{{i}}\,g^{{h}{j}}
      - \gamma^{{j}}\,g^{{h}{i}}
      \big)\,{u}^\tau({p},\chi)\,
\nonumber\\
&& \qquad 
      - \,\tfrac{1}{2}\,v^{}_{2} \left(
          \bar{u}^{{i}}(\bar{p},\bar\chi)\,\gamma^{{j}}\,{u}^{h}({p},\chi)
          - \bar{u}^{{j}}(\bar{p},\bar\chi)\,\gamma^{{i}}\,{u}^{h}({p},\chi)
          + \bar{u}^{{i}}(\bar{p},\bar\chi)\,\gamma^{{h}}\,{u}^{j}({p},\chi)
          - \bar{u}^{{j}}(\bar{p},\bar\chi)\,\gamma^{{h}}\,{u}^{i}({p},\chi)
    \right)\,
\nonumber\\
&& \qquad 
      +\, \tfrac{1}{2}\,v^{}_{3} \left(
          \bar{u}^{{i}}(\bar{p},\bar\chi)\,\gamma^{{h}}\,{u}^{j}({p},\chi)
          - \bar{u}^{{j}}(\bar{p},\bar\chi)\,\gamma^{{h}}\,{u}^{i}({p},\chi)
          + \bar{u}^{{h}}(\bar{p},\bar\chi)\,\gamma^{{i}}\,{u}^{j}({p},\chi)
          - \bar{u}^{{h}}(\bar{p},\bar\chi)\,\gamma^{{j}}\,{u}^{i}({p},\chi)
        \right)\,
    \Big\}\,
  \nonumber\\ && \,
  -\,\tfrac{1}{8}\,\bar{u}_{\tau}(\bar{p},\bar\chi)\,\left(
    i\sigma^{{i}{h}}\,(\bar{p}+{p})^{j} + i\sigma^{{j}{h}}\,(\bar{p}+{p})^{i}
  \right)\,
  {u}^\tau({p},\chi)\,
   \,\Big\{
       (v^{}_{5} + \tfrac{3}{4}\,v^{}_{6})\,d_{abg}\,if_{efg}\,
      \delta^{nop}_{xyz}\,\Lambda^{f,xyz}_{klm}
\nonumber\\
&& \qquad 
      - \,\tfrac{3}{8}\,v^{}_{6}\,
      \delta^{nop}_{rst}\,\big(
      (\Lambda^{a,rst}_{xyz}\,if_{beg} + \Lambda^{b,rst}_{xyz}\,if_{aeg})\,\Lambda^{g,xyz}_{klm}
      + \Lambda^{g,rst}_{xyz}\,(\Lambda^{a,xyz}_{klm}\,if_{beg} + \Lambda^{b,xyz}_{klm}\,if_{aeg})
      \big)
    \Big\}
  \nonumber\\ && \,
  -\, \tfrac{1}{8}\,\bar{u}_{\tau}(\bar{p},\bar\chi)\,\left(
    2\,i\sigma^{{i}{j}}\,(\bar{p}+{p})^{h}
    + i\sigma^{{i}{h}}\,(\bar{p}+{p})^{j} - i\sigma^{{j}{h}}\,(\bar{p}+{p})^{i}
  \right)\,
  {u}^\tau({p},\chi)\,\Big\{
  \nonumber\\
&& \qquad +\,
       2\,(
      \tfrac{2}{3}\,v^{}_{5} + \tfrac{1}{2}\,v^{}_{6} + v^{}_{7}
      )\,\delta^{nop}_{klm}\,if_{abe}
      + (v^{}_{5} + \tfrac{3}{4}\,v^{}_{6})\,(if_{beg}\,d_{agf} - if_{aeg}\,d_{bgf})\,
      \delta^{nop}_{xyz}\,\Lambda^{f,xyz}_{klm}
\nonumber\\
&& \qquad
      - \tfrac{3}{8}\,v^{}_{6}\,
      \delta^{nop}_{rst}\,\big(
      (\Lambda^{a,rst}_{xyz}\,if_{beg} - \Lambda^{b,rst}_{xyz}\,if_{aeg})\,\Lambda^{g,xyz}_{klm}
      + \Lambda^{g,rst}_{xyz}\,(\Lambda^{a,xyz}_{klm}\,if_{beg} - \Lambda^{b,xyz}_{klm}\,if_{aeg})
      \big)
    \Big\}
  \nonumber\\ && \,
  - \, \tfrac{1}{8}\,\delta^{nop}_{xyz}\,\Lambda^{f,xyz}_{klm}\,\big(
  3\,d_{abg}\,d_{feg}
  + \delta_{ab}\,\delta_{fe}
  - \delta_{af}\,\delta_{be} - \delta_{ae}\,\delta_{bf}
  \big)\,
 \Big\{
 \nonumber\\
&& \qquad +\,
     \tfrac{1}{2}\,v_{8}\, \big(
      (
      \bar{u}^{h}(\bar{p},\bar\chi)\,{u}^{j}({p},\chi)
      + \bar{u}^{j}(\bar{p},\bar\chi)\,{u}^{h}({p},\chi)
      )\,(\bar{p}+{p})^{i}
 \nonumber\\
&& \qquad\qquad +\,
      (
      \bar{u}^{h}(\bar{p},\bar\chi)\,{u}^{i}({p},\chi)
      + \bar{u}^{i}(\bar{p},\bar\chi)\,{u}^{h}({p},\chi)
      )\,(\bar{p}+{p})^{j}
      \big)\, 
    \Big\}
  \nonumber\\ &&\,
  - \, \tfrac{1}{8}\,\delta^{nop}_{xyz}\,\Lambda^{f,xyz}_{klm}\,
  f_{abg}\,f_{feg}\,
 \Big\{
 \nonumber\\
&& \qquad +\,
     \tfrac{1}{2}\,v_{8}\,  \big(
      (
      \bar{u}^{h}(\bar{p},\bar\chi)\,{u}^{j}({p},\chi)
      + \bar{u}^{j}(\bar{p},\bar\chi)\,{u}^{h}({p},\chi)
      )\,(\bar{p}+{p})^{i}
 \nonumber\\
&& \qquad\qquad -\,
      (
      \bar{u}^{h}(\bar{p},\bar\chi)\,{u}^{i}({p},\chi)
      + \bar{u}^{i}(\bar{p},\bar\chi)\,{u}^{h}({p},\chi)
      )\,(\bar{p}+{p})^{j}
      \big)\,
    \Big\}
  \nonumber\\ && \,
  +\, \tfrac{1}{4}\, \delta^{nop}_{xyz}\,\Lambda^{d,xyz}_{klm}\,
  \big( 3\,d_{abg}\,d_{edg}
      + \delta_{ab}\,\delta_{ed}
      - \delta_{ae}\,\delta_{bd} - \delta_{ad}\,\delta_{be} \big)
  \,\Big\{
\nonumber\\
&& \qquad
    + v^{}_{4}\,\bar{u}_\tau(\bar{p},\bar\chi)\,\big(
    \bar{p}^{i}\,\gamma^{j}
    + \bar{p}^{j}\,\gamma^{i}\,
  \big)\,{u}^\tau({p},\chi)\,(\bar{p}+p)^{h}
    \Big\}
   \nonumber\\ && \,
   - \,  \tfrac{1}{4}\,   \bar{u}_\tau(\bar{p},\bar\chi)\,\big(
      \, \bar{p}^{i}\,\gamma^{j}\,
       - \bar{p}^{j}\,\gamma^{i}
     \big)\,{u}^\tau({p},\chi)\,(\bar{p}-p)^{h}
   \,\delta^{nop}_{xyz}\,\Lambda^{d,xyz}_{klm}\,
    v^{}_{4}\, f_{abg}\,f_{edg} \,,
\end{eqnarray}
and
\clearpage
\begin{eqnarray}
  && \qquad \qquad \qquad \qquad \qquad \qquad
  \bra{\bar{p},\,\bar \chi,\,nop\, }\, {\mathcal O}^{(abe)}_{ijh}(q,q') \,\ket{ p, \chi, c}
  \nonumber\\ \nonumber\\
  && = \,
  \tfrac{1}{8\,\sqrt{2}} \,\big(
  \bar{u}^{i}(\bar{p},\bar\chi)\,i\,\sigma^{{j}{h}} + \bar{u}^{j}(\bar{p},\bar\chi)\,i\,\sigma^{{i}{h}}
  \big)\,\gamma_5\,{u}({p},\chi)\,\Big\{
      - w^{}_{1}\,d_{abf}\,i\,f_{egf}
 \nonumber\\
&& \qquad      
      -\, w^{}_{4}\,\big(3\,d_{abf}\,d_{gef}
      + \delta_{ab}\,\delta_{ge} - (\delta_{ag}\,\delta_{be} + \delta_{bg}\,\delta_{ae})\big)
    \Big\}\,\Lambda^{nop}_{gc}
  \nonumber\\ && \,
  + \,\tfrac{1}{16\,\sqrt{2}} \,\big(
   ( \bar{u}^{i}(\bar{p},\bar\chi)\,\gamma^{j} + \bar{u}^{j}(\bar{p},\bar\chi)\,\gamma^{i}
  )\,(\bar{p}+{p})^{h}
 \nonumber\\
&& \qquad \qquad \qquad  
  +(
    \bar{u}^{i}(\bar{p},\bar\chi)\,(\bar{p}+{p})^{j} + \bar{u}^{j}(\bar{p},\bar\chi)\,(\bar{p}+{p})^{i}
  )\,\gamma^{h}
  \big)\,\gamma_5\,{u}({p},\chi)\,\Big\{
      - w^{}_{2}\,d_{abf}\,i\,f_{egf}
 \nonumber\\
&& \qquad
      - w^{}_{5}\,\big(3\,d_{abf}\,d_{gef}
      + \delta_{ab}\,\delta_{ge} - (\delta_{ag}\,\delta_{be} + \delta_{bg}\,\delta_{ae})\big)
    \Big\}\,\Lambda^{nop}_{gc}
  \nonumber\\ && \,
  +\,\tfrac{1}{16\,\sqrt{2}} \,\big(
  (
    \bar{u}^{i}(\bar{p},\bar\chi)\,\gamma^{j} + \bar{u}^{j}(\bar{p},\bar\chi)\,\gamma^{i}
  )\,(\bar{p}+{p})^{h} 
   \nonumber\\
&& \qquad \qquad \qquad 
  +\,\bar{u}^{h}(\bar{p},\bar\chi)\,(
    \gamma^{i}\,(\bar{p}+{p})^{j} + \gamma^{j}\,(\bar{p}+{p})^{i}
  )
  \big)\,\gamma_5\,{u}({p},\chi)\,\Big\{
      - w^{}_{3}\,d_{abf}\,if_{egf}
 \nonumber\\
&& \qquad
      - w^{}_{6}\,\big(3\,d_{abf}\,d_{gef}
      + \delta_{ab}\,\delta_{ge} - (\delta_{ag}\,\delta_{be} + \delta_{bg}\,\delta_{ae})\big)
    \Big\}\,\Lambda^{nop}_{gc}
  \nonumber\\ && \,
  +\tfrac{1}{8\,\sqrt{2}}\, \big(
 2\, \bar{u}^{h}(\bar{p},\bar\chi)\,i\,\sigma^{{i}{j}}
  - (\bar{u}^{i}(\bar{p},\bar\chi)\,i\,\sigma^{{j}{h}} - \bar{u}^{j}(\bar{p},\bar\chi)\,i\,\sigma^{{i}{h}})
  \big)\,\gamma_5\,{u}({p},\chi)
  \,\Big\{
        w^{}_{1}\,\big(if_{aef}\,d_{bgf} - if_{bef}\,d_{agf}\big)
 \nonumber\\
&& \qquad
      - w^{}_{4}\,f_{abf}\,f_{gef}
    \Big\}\,\Lambda^{nop}_{gc}
  \nonumber\\ && \,
  +\,\tfrac{1}{16\,\sqrt{2}} \,\big(
  (
    \bar{u}^{i}(\bar{p},\bar\chi)\,\gamma^{j} - \bar{u}^{j}(\bar{p},\bar\chi)\,\gamma^{i}
  )\,(\bar{p}+{p})^{h}
   \nonumber\\
&& \qquad \qquad \qquad 
  +\,(
    \bar{u}^{i}(\bar{p},\bar\chi)\,(\bar{p}+{p})^{j} - \bar{u}^{j}(\bar{p},\bar\chi)\,(\bar{p}+{p})^{i}
  )\,\gamma^{h}
  \big)\,\gamma_5\,{u}({p},\chi)\,\Big\{
      w^{}_{2}\,\big(i\,f_{aef}\,d_{bgf} - i\,f_{bef}\,d_{agf}\big)
 \nonumber\\
&& \qquad
            - w^{}_{5}\,f_{abf}\,f_{gef}
    \Big\}\,\Lambda^{nop}_{gc}
  \nonumber\\ && \,
  +\,\tfrac{1}{16\,\sqrt{2}} \,\big(  
  -\,(
    \bar{u}^{i}(\bar{p},\bar\chi)\,\gamma^{j} - \bar{u}^{j}(\bar{p},\bar\chi)\,\gamma^{i}
  )\,(\bar{p}+{p})^{h} 
   \nonumber\\
&& \qquad \qquad \qquad 
  +\,\bar{u}^{h}(\bar{p},\bar\chi)\,(
    \gamma^{i}\,(\bar{p}+{p})^{j} - \gamma^{j}\,(\bar{p}+{p})^{i}
  )
  \big)\,\gamma_5\,{u}({p},\chi)\,\Big\{
      w^{}_{3}\,\big(if_{aef}\,d_{bgf} - if_{bef}\,d_{agf}\big)
 \nonumber\\
&& \qquad
      - w^{}_{6}\,f_{abf}\,f_{gef}
    \Big\}\,\Lambda^{nop}_{gc}
  \,.
\end{eqnarray}

\clearpage

\section{Appendix B}
\begin{eqnarray}
  \roundbra{{d}, {\bar\chi}}\,
  [[T_a,\,T_b]_+,\,T_e]_+
  \,\roundket{{c}, \chi}
  & = &
  \delta_{\bar\chi\chi}\,\big[
  2\,\delta_{ab}\,if_{e{c}{d}}
  + 3\,d_{abg}\,(
  d_{cgf}\,if_{{d}ef}
  - d_{{d}gf}\,if_{{c}ef}
  )
  \nonumber\\ && \qquad
  - \delta_{b{d}}\,if_{e{c}a}
  - \delta_{a{d}}\,if_{e{c}b}
  - \delta_{ac}\,if_{eb{d}}
  - \delta_{bc}\,if_{ea{d}}
  \big]
  \,,\nonumber\\
  \roundbra{{d}, {\bar\chi}}\,
  [[G^i_a,\,G^j_b]_+,\,T_e]_+
  \,\roundket{{c}, \chi}
  & = &
  \tfrac{1}{4}\,\delta_{\bar\chi\chi}\,\delta^{ij}\,\big[
  \tfrac{10}{3}\,\delta_{ab}\,if_{e{c}{d}}
  \nonumber\\ && \qquad
  - d_{abg}\,(
  d_{g{d}f}\,if_{e{c}f}
  - d_{gcf}\,if_{e{d}f}
  )
  + \tfrac{8}{3}\,d_{abg}\,(
  f_{g{d}f}\,f_{e{c}f}
  + f_{gcf}\,f_{e{d}f}
  )
  \nonumber\\ && \qquad
  + \tfrac{1}{3}\,(
  \delta_{a{d}}\,if_{eb{c}}
  + \delta_{b{d}}\,if_{ea{c}}
  - \delta_{ac}\,if_{eb{d}}
  - \delta_{bc}\,if_{ea{d}}
  )
  \big]
  \nonumber\\ &+&
  \tfrac{1}{4}\,i\,\epsilon^{ijk}\,\sigma^k_{\bar\chi\chi}\,\big[
  \delta_{a{d}}\,if_{eb{c}}
  - \delta_{b{d}}\,if_{ea{c}}
  + \delta_{ac}\,if_{eb{d}}
  - \delta_{bc}\,if_{ea{d}}
  \nonumber\\ && \qquad
  + 2\,if_{abg}\,(
  d_{g{d}f}\,if_{e{c}f}
  - d_{gcf}\,if_{e{d}f}
  )
  \nonumber\\ && \qquad
  + \tfrac{5}{3}\,if_{abg}\,(
  f_{g{d}f}\,f_{e{c}f}
  + f_{gcf}\,f_{e{d}f}
  )
  \big]
  \,,\nonumber\\
  \roundbra{{d}, {\bar\chi}}\,
  [[T_a,\,T_b]_+,\,G^i_e]_+
  \,\roundket{{c}, \chi}
  & = &
  \tfrac{1}{2}\,\sigma^i_{\bar\chi\chi}\,\big[
  2\,\delta_{ab}\,d_{ec{d}}
  + \tfrac{4}{3}\,\delta_{ab}\,if_{ec{d}}
  \nonumber\\ && \qquad
  + 3\,d_{abg}\,d_{fg{d}}\,d_{ecf}
  + 2\,d_{abg}\,d_{fg{d}}\,if_{ecf}
  \nonumber\\ && \qquad
  + 3\,d_{abg}\,d_{cgf}\,d_{ef{d}}
  + 2\,d_{abg}\,d_{cgf}\,if_{ef{d}}
  \nonumber\\ && \qquad
  - \delta_{ac}\,d_{eb{d}}
  - \delta_{bc}\,d_{ea{d}}
  - \delta_{b{d}}\,d_{eca}
  - \delta_{a{d}}\,d_{ecb}
  \nonumber\\ && \qquad
  - \tfrac{2}{3}\,\delta_{ac}\,if_{eb{d}}
  - \tfrac{2}{3}\,\delta_{bc}\,if_{ea{d}}
  - \tfrac{2}{3}\,\delta_{b{d}}\,if_{eca}
  - \tfrac{2}{3}\,\delta_{a{d}}\,if_{ecb}
  \big]
  \,,\nonumber\\
  \roundbra{{d}, {\bar\chi}}\,
  [[T_a,\,T_b]_+,\,J^i]_+
  \,\roundket{{c}, \chi}
  & = &
  \sigma^i_{\bar\chi\chi}\,\big[
  \delta_{ab}\,\delta_{{c}{d}} - \delta_{a{c}}\,\delta_{b{d}} - \delta_{a{d}}\,\delta_{b{c}}
  + 3\,d_{abg}\,d_{{c}g{d}}
  \big]
  \,,\nonumber\\
  \roundbra{{d}, {\bar\chi}}\,
  [[G^i_a,\,G^l_b]_+,\,J^l]_+
  \,\roundket{{c}, \chi}
  & = &
  \tfrac{1}{4}\,\sigma^i_{\bar\chi\chi}\,\big[
  \tfrac{1}{3}\,(
  5\,\delta_{ab}\,\delta_{c{d}} - \delta_{ac}\,\delta_{b{d}} - \delta_{a{d}}\,\delta_{bc}
  )
  \nonumber\\ && \qquad
  - d_{{d}cg}\,d_{abg}
  + \tfrac{8}{3}\,i\,f_{c{d}g}\,d_{abg}
  \big]
  \,,\nonumber\\ \nonumber\\\nonumber\\
  \roundbra{{nop}, {\bar\chi}}\,
  [[T_a,\,T_b]_+,\,T_e]_+
  \,\roundket{{klm}, \chi}
  & = &
  \tfrac{27}{8}\,\delta_{\bar\chi\chi}\,\delta^{{nop}}_{rst}\,\Big\{
  \Lambda^{a,rst}_{xyz}\,\Lambda^{b,xyz}_{uvw}\,\Lambda^{e,uvw}_{klm}
  + \Lambda^{e,rst}_{uvw}\,\Lambda^{a,uvw}_{xyz}\,\Lambda^{b,xyz}_{klm}
  +\,(a \leftrightarrow b)
  \Big\}
  \,,\nonumber\\
  \roundbra{{nop}, {\bar\chi}}\,
  [[G^i_a,\,G^j_b]_+,\,T_e]_+
  \,\roundket{{klm}, \chi}
  & = &
  \tfrac{3}{2}\,\delta^{{nop}}_{rst}\,\Big\{
  \nonumber\\ && \,
  \!\!\!\!\!\!\!\!\!\!\!\!\!\!\!\!\!\!\!\!\!\!\!\!\!\!\!\!\!\!\!\!\!\!\!\!\!\!\!\!\!
  \!\!\!\!\!\!\!\!\!\!\!\!\!\!\!\!\!\!\!\!\!\!\!\!\!\!\!\!\!\!\!\!\!\!\!
  - \tfrac{3}{8}\,(S^{i}\,S^{j\dagger} + S^{j}\,S^{i\dagger} - \tfrac{3}{2}\,\delta^{ij}\,\one_{(4\times 4)}
  )_{\bar\chi\chi}\,\big[
  \Lambda^{a,rst}_{xyz}\,\Lambda^{b,xyz}_{uvw}\,\Lambda^{e,uvw}_{klm}
  + \Lambda^{e,rst}_{uvw}\,\Lambda^{a,uvw}_{xyz}\,\Lambda^{b,xyz}_{klm}
  +\,(a \leftrightarrow b)
  \big]
  \nonumber\\ && \,
  \!\!\!\!\!\!\!\!\!\!\!\!\!\!\!\!\!\!\!\!\!\!\!\!\!\!\!\!\!\!\!\!\!\!\!\!\!\!\!\!\!
  \!\!\!\!\!\!\!\!\!\!\!\!\!\!\!\!\!\!\!\!\!\!\!\!\!\!\!\!\!\!\!\!\!\!\!
  + \tfrac{3}{16}\,i\,\epsilon^{ijk'}\,(\vec{S}\,\sigma^{k\prime}\,\vec{S}^\dagger
  )_{\bar\chi\chi}\,\big[
  \Lambda^{a,rst}_{xyz}\,\Lambda^{b,xyz}_{uvw}\,\Lambda^{e,uvw}_{klm}
  + \Lambda^{e,rst}_{uvw}\,\Lambda^{a,uvw}_{xyz}\,\Lambda^{b,xyz}_{klm}
  -\,(a \leftrightarrow b)
  \big]
  \nonumber\\ && \,
  \!\!\!\!\!\!\!\!\!\!\!\!\!\!\!\!\!\!\!\!\!\!\!\!\!\!\!\!\!\!\!\!\!\!\!\!\!\!\!\!\!
  \!\!\!\!\!\!\!\!\!\!\!\!\!\!\!\!\!\!\!\!\!\!\!\!\!\!\!\!\!\!\!\!\!\!\!
  + \tfrac{1}{16}\,(S^{i}\,S^{j\dagger} + S^{j}\,S^{i\dagger}
  )_{\bar\chi\chi}\,\big[
  \Lambda^{rst}_{ag}\,\Lambda^{bg}_{uvw}\,\Lambda^{e,uvw}_{klm}
  + \Lambda^{e,rst}_{uvw}\,\Lambda^{uvw}_{ag}\,\Lambda^{bg}_{klm}
  +\,(a \leftrightarrow b)
  \big]
  \nonumber\\ && \,
  \!\!\!\!\!\!\!\!\!\!\!\!\!\!\!\!\!\!\!\!\!\!\!\!\!\!\!\!\!\!\!\!\!\!\!\!\!\!\!\!\!
  \!\!\!\!\!\!\!\!\!\!\!\!\!\!\!\!\!\!\!\!\!\!\!\!\!\!\!\!\!\!\!\!\!\!\!
  + \tfrac{1}{16}\,i\,\epsilon^{ijk'}\,(\vec{S}\,\sigma^{k\prime}\,\vec{S}^\dagger
  )_{\bar\chi\chi}\,\big[
  \Lambda^{rst}_{ag}\,\Lambda^{bg}_{uvw}\,\Lambda^{e,uvw}_{klm}
  + \Lambda^{e,rst}_{uvw}\,\Lambda^{uvw}_{ag}\,\Lambda^{bg}_{klm}
  -\,(a \leftrightarrow b)
  \big]
  \Big\}
  \,,\nonumber\\
  \roundbra{{nop}, {\bar\chi}}\,
  [[T_a,\,T_b]_+,\,G^i_e]_+
  \,\roundket{{klm}, \chi}
  & = &
  \tfrac{27}{16}\,(\vec{S}\,\sigma^i\,\vec{S}^\dagger)_{\bar\chi\chi}\,\delta^{{nop}}_{rst}\,
  \Big\{
  \nonumber\\ && \qquad
  \Lambda^{a,rst}_{xyz}\,\Lambda^{b,xyz}_{uvw}\,\Lambda^{e,uvw}_{klm}
  + \Lambda^{e,rst}_{xyz}\,\Lambda^{a,xyz}_{uvw}\,\Lambda^{b,uvw}_{klm}
  +\,(a \leftrightarrow b)
  \Big\}
  \,,\nonumber\\
  \roundbra{{nop}, {\bar\chi}}\,
  [[T_a,\,T_b]_+,\,J^i]_+
  \,\roundket{{klm}, \chi}
  & = &
  \tfrac{27}{4}\,(\vec{S}\,\sigma^i\,\vec{S}^\dagger)_{\bar\chi\chi}\,
  \delta^{{nop}}_{rst}\,\Big\{
  \Lambda^{a,rst}_{xyz}\,\Lambda^{b,xyz}_{klm}
  +\,(a \leftrightarrow b)
  \Big\}
  \,,\nonumber\\
  \roundbra{{nop}, {\bar\chi}}\,
  [[G^i_a,\,G^l_b]_+,\,J^l]_+
  \,\roundket{{klm}, \chi}
  & = &
  \tfrac{3}{2}\,(\vec{S}\,\sigma^i\,\vec{S}^\dagger)_{\bar\chi\chi}\,
  \delta^{{nop}}_{rst}\,\Big\{
  \nonumber\\ && \qquad
  \tfrac{13}{8}\,\,\Lambda^{a,rst}_{xyz}\,\Lambda^{b,xyz}_{klm}
  - \tfrac{1}{12}\,\Lambda^{rst}_{ag}\,\Lambda^{bg}_{klm}
  +\,(a \leftrightarrow b)
  \Big\}
  \Big\}
  \,,\nonumber\\ \nonumber\\\nonumber\\
  \roundbra{{nop}, {\bar\chi}}\,
  [[T_a,\,T_b]_+,\,T_e]_+
  \,\roundket{{c}, \chi}
  & = &
  0
  \,,\nonumber\\
  \roundbra{{nop}, {\bar\chi}}\,
  [[G^i_a,\,G^j_b]_+,\,T_e]_+
  \,\roundket{{c}, \chi}
  & = &
  \tfrac{1}{8\,\sqrt{2}}\,\delta^{{nop}}_{rst}\,\Big\{
  \nonumber\\ && \,
  \!\!\!\!\!\!\!\!\!\!\!\!\!\!\!\!\!\!\!\!\!\!\!\!\!\!\!\!\!\!\!\!\!\!\!\!\!\!\!\!\!
  \!\!\!\!\!\!\!\!\!\!\!\!\!\!\!\!\!\!\!\!\!\!\!\!\!\!\!\!\!\!\!\!\!\!\!
  (S^i\,\sigma^j+S^j\,\sigma^i)_{\bar\chi\chi}\,\big[
  if_{e{c}f}\,(d_{afg}+i\,f_{afg})\,\Lambda^{rst}_{bg}
  + \tfrac{3}{2}\,\Lambda^{e,rst}_{uvw}\,(d_{acg}+i\,f_{acg})\,\Lambda^{uvw}_{bg}
  +\,(a \leftrightarrow b)
  \big]
  \nonumber\\ && \,
  \!\!\!\!\!\!\!\!\!\!\!\!\!\!\!\!\!\!\!\!\!\!\!\!\!\!\!\!\!\!\!\!\!\!\!\!\!\!\!\!\!
  \!\!\!\!\!\!\!\!\!\!\!\!\!\!\!\!\!\!\!\!\!\!\!\!\!\!\!\!\!\!\!\!\!\!\!
  + i\,\epsilon^{ijk}\,S^k_{\bar\chi\chi}\,\big[
  if_{e{c}f}\,\big(
  (d_{afg}+\tfrac{2}{3}\,i\,f_{afg})\,\Lambda^{rst}_{bg}
  + \tfrac{5}{3}\,(
  i\,f_{afg}\,\Lambda^{rst}_{bg}
  - i\,f_{abg}\,\Lambda^{rst}_{fg}
  )
  \big)
  \nonumber\\ &&
  \!\!\!\!\!\!\!\!\!\!\!\!\!\!\!\!\!\!\!\!\!\!\!\!\!\!\!\!\!\!\!\!\!\!\!
  \!\!\!\!\!\!\!\!\!\!\!\!
  + \tfrac{3}{2}\,\Lambda^{e,rst}_{uvw}\,\big(
  (d_{acg}+\tfrac{2}{3}\,i\,f_{acg})\,\Lambda^{uvw}_{bg}
  + \tfrac{5}{3}\,(
  i\,f_{acg}\,\Lambda^{uvw}_{bg}
  - i\,f_{abg}\,\Lambda^{uvw}_{cg}
  )
  \big)
  -\,(a \leftrightarrow b)
  \big]
  \Big\}
  \,,\nonumber\\
  \roundbra{{nop}, {\bar\chi}}\,
  [[T_a,\,T_b]_+,\,G^i_e]_+
  \,\roundket{{c}, \chi}
  & = &
  \tfrac{1}{2\sqrt{2}}\,S^i_{\bar\chi\chi}\,\Big\{
  \tfrac{9}{4}\,\delta^{{nop}}_{rst}\,(
  \Lambda^{a,rst}_{xyz}\,\Lambda^{b,xyz}_{uvw}
  + \Lambda^{b,rst}_{xyz}\,\Lambda^{a,xyz}_{uvw}
  )\,\Lambda^{{uvw}}_{ec}
  \nonumber\\ && \qquad
  + \Lambda^{{nop}}_{ef}\big(
  \delta_{ab}\,\delta_{cf} - \delta_{ac}\,\delta_{bf} - \delta_{af}\,\delta_{bc}
  + 3\,d_{abg}\,d_{cgf}
  \big)
  \Big\}
  \,,\nonumber\\
  \roundbra{{nop}, {\bar\chi}}\,
  [[T_a,\,T_b]_+,\,J^i]_+
  \,\roundket{{c}, \chi}
  & = &
  0
  \,,\nonumber\\
  \roundbra{{nop}, {\bar\chi}}\,
  [[G^i_a,\,G^l_b]_+,\,J^l]_+
  \,\roundket{{c}, \chi}
  & = &
  \tfrac{1}{8\,\sqrt{2}}\,S^i_{\bar\chi\chi}\,\Big\{
  5\,(d_{a{c}g}+i\,f_{a{c}g})\,\Lambda^{{nop}}_{bg}
  + 5\,(d_{b{c}g}+i\,f_{b{c}g})\,\Lambda^{{nop}}_{ag}
  \nonumber\\ && \qquad
  - 3\,(d_{a{c}g}+\tfrac{2}{3}\,i\,f_{a{c}g})\,\Lambda^{{nop}}_{bg}
  + 3\,(d_{b{c}g}+\tfrac{2}{3}\,i\,f_{b{c}g})\,\Lambda^{{nop}}_{ag}
  \nonumber\\ && \qquad
  - 5\,(
  i\,f_{a{c}g}\,\Lambda^{{nop}}_{bg}
  - i\,f_{b{c}g}\,\Lambda^{{nop}}_{ag}
  - 2\,i\,f_{abg}\,\Lambda^{{nop}}_{{c}g}
  )
  \Big\}
  \label{res-three-body}
  \,,
\end{eqnarray}
where Eq.~(A.2) in~\cite{Lutz:2010se} was used. We correct a typo with
\begin{eqnarray}
  \roundbra{{nop}, {\bar\chi}}\,[G^i_a,G^j_b]_+\,\roundket{{c}, \chi}
  & = &
  \tfrac{1}{8\,\sqrt{2}}\,i\,\epsilon^{ijk}\,S^k_{\bar\chi\chi}\,\big[
  (d_{ace}+\tfrac{2}{3}\,i\,f_{ace})\,\Lambda^{{nop}}_{be}
  + \tfrac{5}{3}\,(
  i\,f_{ace}\,\Lambda^{{nop}}_{be}
  - i\,f_{abe}\,\Lambda^{{nop}}_{ce}
  )
  -\,(a \leftrightarrow b)
  \big]
  \nonumber\\ &+&
  \tfrac{1}{8\,\sqrt{2}}\,(S^i\,\sigma^j+S^j\,\sigma^i)_{\bar\chi\chi}\,\big[
  (d_{ace}+i\,f_{ace})\,\Lambda^{{nop}}_{be}
  +\,(a \leftrightarrow b)
  \big]
  \,.
\end{eqnarray}

\newpage
\bibliography{literature}
\end{document}